\documentclass[11pt,a4paper]{article}

\usepackage{verbatim}
\usepackage{graphicx}
\usepackage{graphics}
\usepackage[footnotesize,bf,hang]{caption}
\usepackage{natbib}
\usepackage[dvips]{color}
\usepackage{amsmath}
\usepackage{amssymb}
\usepackage{vmargin}
\usepackage[symbol]{footmisc}

\setmarginsrb{35mm}{40mm}{30mm}{40mm}{0pt}{0mm}{0pt}{5mm}

\begin{document}
\begin{center} 
\begin{figure*}[ht]
{\small
\framebox[6in]{
\begin{minipage}[t]{5.5in}
\emph{The definitive version of this manuscript is available in the Journal of the Royal Statistical Society, Series C, at www3.interscience.wiley.com/journal/117997424/home, 2010, Vol. 59, Part 2, p. 255-277.}
\end{minipage}
}
}
\end{figure*}
\LARGE \textbf{Estimating infectious disease parameters from data on social contacts and serological status} \\ \vspace{1cm}
\large \textbf{Nele Goeyvaerts$^{1}$\footnote{Address for correspondence: Nele Goeyvaerts, Hasselt University, Campus Diepenbeek, Department WNI /TWI, Agoralaan 1 Gebouw D, B-3590 Diepenbeek, Belgium. \\ Email: nele.goeyvaerts@uhasselt.be}, Niel Hens$^{1,2}$, Benson Ogunjimi$^2$, Marc Aerts$^1$, Ziv Shkedy$^1$, Pierre Van Damme$^2$, Philippe Beutels$^2$} \\ \vspace{1cm} \small
$^1$Interuniversity Institute for Biostatistics and Statistical Bioinformatics, Hasselt University, Agoralaan 1, B3590 Diepenbeek, Belgium \\
$^2$Centre for Health Economics Research and Modeling Infectious Diseases (CHERMID) \& Centre for the Evaluation of Vaccination (CEV), Vaccine \& Infectious Disease Institute, University of Antwerp, Belgium
\end{center}

\normalsize

\begin{abstract}
In dynamic models of infectious disease transmission, typically various mixing patterns are imposed on the so-called Who-Acquires-Infection-From-Whom matrix (WAIFW). These imposed mixing patterns are based on prior knowledge of age-related social mixing behavior rather than observations. Alternatively, one can assume that transmission rates for infections transmitted predominantly through non-sexual social contacts, are proportional to rates of conversational contact which can be estimated from a contact survey. In general, however, contacts reported in social contact surveys are proxies of those events by which transmission may occur and there may exist age-specific characteristics related to susceptibility and infectiousness which are not captured by the contact rates. Therefore, in this paper, transmission is modeled as the product of two age-specific variables: the age-specific contact rate and an age-specific proportionality factor, which entails an improvement of fit for the seroprevalence of the varicella-zoster virus (VZV) in Belgium. Furthermore, we address the impact on the estimation of the basic reproduction number, using non-parametric bootstrapping to account for different sources of variability and using multi-model inference to deal with model selection uncertainty. The proposed method makes it possible to obtain important information on transmission dynamics that cannot be inferred from approaches traditionally applied hitherto.

\end{abstract}

\section{Introduction}

A first approach in modeling transmission dynamics of infectious diseases, and more particularly in estimating age-dependent transmission rates, was described by \cite{Anderson1991}.  The idea is to impose different mixing patterns on the so-called WAIFW-matrix $\beta_{ij}$, hereby constraining the number of distinct elements for identifiability reasons, and to estimate the parameters from serological data. Many authors have elaborated on this approach of \cite{Anderson1991}, among which \cite{Greenhalgh1994}, \cite{Farrington2001} and \cite{Van2009}. However, estimates of important epidemiological parameters such as the basic reproduction number $R_0$ turn out to be sensitive with respect to the choice of the imposed mixing pattern \citep{Greenhalgh1994}.

An alternative method was proposed by \cite{Farrington2005}, where contact rates are modeled as a continuous contact surface and estimated from serological data.
Clearly, both methods involve a somewhat ad hoc choice, namely the structure for the WAIFW-matrix and the parametric model for the contact surface.
Alternatively, to estimate age-dependent transmission parameters, \cite{Wallinga2006} augmented seroprevalence data with auxiliary data on self-reported numbers of conversational contacts per person, whilst assuming that transmission rates are proportional to rates of conversational contact.
The social contact surveys conducted as part of the POLYMOD project \citep{Mossong2008b,Hens2009a}, allow us to elaborate on this methodology presented by \cite{Wallinga2006}.

The paper is organized as follows. In the next section, we outline the buildup of the Belgian social contact survey and the information available for each contact. Further, we briefly explain the epidemiological characteristics of VZV and the serological data from Belgium we use.  In Section 3, we illustrate the traditional approach of imposing mixing patterns to estimate the WAIFW-matrix from this serological data set. In Section 4, a transition is made to the novel approach of using social contact data to estimate $R_0$. We show that a bivariate smoothing approach allows for a more flexible and better estimate of the contact surface compared to the maximum likelihood estimation method of \cite{Wallinga2006}. Further, some refinements are proposed, among which an elicitation of contacts with high transmission potential and a non-parametric bootstrap approach, assessing sampling variability and accounting for age uncertainty, as suggested by \cite{Halloran2006}.

Our main result is the novel method of disentangling the WAIFW-matrix into two components: the contact surface and an age-dependent proportionality factor. The proposed method, as described in Section 5, tackles two dimensions of uncertainty. First, by estimating the contact surface from data on social contacts, we overcome the problem of choosing a completely parametric model for the WAIFW-matrix.  Second, to overcome the problem of model selection for the age-dependent proportionality factor, concepts of multi-model inference are applied and a model averaged estimate for $R_0$ is calculated.  Some concluding remarks are provided in the last section.

\section{Data}

\subsection{Belgian contact survey} \label{contact data}

Several small scale surveys were made in order to gain more insight in social mixing behavior relevant to the spread of close contact infections \citep{Edmunds1997,Beutels2006,Edmunds2006,Wallinga2006,Mikolajczyk2008}. In order to refine on contact information, a large multi-country population-based survey was conducted in Europe as part of the POLYMOD project \citep{Mossong2008b}.

In Belgium, this survey was conducted in a period from March until May 2006. A total of 750 participants, selected through random digit dialing, completed a diary-based questionnaire about their social contacts  during one randomly assigned weekday and one randomly assigned day in the weekend (not always in that order). In this paper, we follow the sampling scheme of the POLYMOD project and only consider one day for each participant \citep{Mossong2008b}.  The data set consists of participant-related information such as age and gender, and details about each contact: age and gender of the contacted person, and location, duration and frequency of the contact. In case the exact age of the contacted person was unknown, participants had to provide an estimated age range and the mean value is used as a surrogate. Further, a distinction between two types of contacts was made: non-close contacts, defined as two-way conversations of at least three words in each others proximity, and close contacts that involve any sort of physical skin-to-skin touching.

Teenagers (9-17y) filled in a simplified version of the diary and were closely followed up to anticipate interpretation problems. For children ($<9$y), a parent or exceptionally another adult caregiver filled in the diary.  One adult respondent made over 1000 contacts and was considered an outlier to the data set. This person is likely very influential and therefore excluded from the analyses presented here. Analyses are based on the remaining 749 participants. Using census data on population sizes of different age by household size combinations, weights are given to the participants in order to make the data representative of the Belgian population. In total, the 749 participants recorded 12775 contacts of which 3 are omitted from analysis due to missing age values for the contacted person. For a more in depth perspective on the Belgian contact survey and the importance of contact rates on modeling infectious diseases, we refer to \cite{Hens2009a}.

\subsection{Serological data}

Primary infection with VZV, also known as human herpes virus 3 (HHV-3), results in varicella, commonly known as chickenpox, and mainly occurs in childhood. Afterwards, the virus becomes dormant in the body and may reactivate in a later stage, resulting in herpes zoster, commonly known as shingles. Infection with VZV occurs through direct or aerosol contact with infected persons. A person infected with chickenpox is able to transmit the virus for about 7 days. Following \cite{Garnett1992} and \cite{Whitaker2004}, we ignore chickenpox cases resulting from contact with persons suffering from shingles. Zoster indeed has a limited impact on transmission dynamics when considering large populations with no immunization program \citep{Ferguson1996}.

In a period from November 2001 until March 2003, 2655 serum samples in Belgium were collected and tested for VZV. Together with the test results, gender and age of the individuals were recorded. In the data set, age ranges from 0 to 40 years and 6 individuals are younger than 6 months. Belgium has no mass vaccination program for VZV. Further details on the data set can be found in \cite{Hens2008} and \cite{Hens2009b}.

\section{Estimation of $R_0$ by imposing mixing patterns}

\subsection{Estimating transmission rates}

To describe transmission dynamics, a compartmental MSIR-model for a closed population of size $N$ is considered. By doing so, we explicitly take into account the fact that, in a first phase, newborns are protected by maternal antibodies and do not take part in the transmission process. We assume that mortality due to infection can be ignored, which is plausible for VZV in developed countries, and that infected individuals maintain lifelong immunity after recovery. Further, demographic and endemic equilibrium are assumed, which means that the age-specific population sizes remain constant over time and that the disease is in an endemic steady state at the population level.
For simplicity, we assume type I mortality defined as
\[ \exp \left(-\int_{0}^a \mu(s) ds \right) = \left\{\begin{array}{ll} 1, &\mbox{if } a < L \\
0, &\mbox{if } a \geq L,
\end{array}\right. \]
where $\mu(a)$ denotes the age-specific mortality rate. This implies that everyone survives up to age $L$ and then promptly dies, which is a reasonable assumption when describing transmission dynamics for VZV in Belgium \citep[see also][]{Whitaker2004}. We make a similar assumption for the age-specific rate $\gamma(a)$ of losing maternal antibodies, which we will denote as `type I maternal antibodies':
\begin{eqnarray}
\exp \left(-\int_{0}^a \gamma(s) ds \right) = \left\{\begin{array}{ll} 1, &\mbox{if } a \leq  A \\
0, &\mbox{if } a > A,
\end{array}\right.
\label{eq:MA}
\end{eqnarray}
meaning that all newborns are protected by maternal antibodies until a certain age $A$ and then move to the susceptible class instantaneously.
Under these assumptions, the proportion of susceptibles is given by
\begin{eqnarray} \label{eq:susc}
x(a) = \exp \left(-\int_{A}^a \lambda(s) ds \right),  ~~~~\mbox{if } a > A,
\end{eqnarray}
where $\lambda(a)$ denotes the age-specific force of infection, and $x(a)=0$ if $a \leq A$.

If the mean duration of infectiousness $D$ is short compared to the timescale on which transmission and mortality rate vary, the force of infection can be approximated by \citep{Anderson1991}:
\begin{equation}
\lambda(a) = \frac{ND}{L} \int_A^{\infty} \beta(a,a') \lambda(a') x(a') da',
\label{foi}
\end{equation}
where $\beta(a,a')$ denotes the transmission rate i.e. the per capita rate at which an individual of age $a'$ makes an effective contact with a person of age $a$, per year. Formula (\ref{foi}) reflects the so-called `mass action principle', which implicitly assumes that infectious and susceptible individuals mix completely with each other and move randomly within the population.

Estimating transmission rates using seroprevalence data can not be done analytically since the integral equation (\ref{foi}) in general has no closed form solution. However, it is possible to solve this numerically by turning to a discrete age framework, assuming a constant force of infection in each age-class.  Denote the first age interval $(a_{[1]},a_{[2]})$ and the $j{\mbox{th}}$ age interval $[a_{[j]},a_{[j+1]})$, $j=2,\ldots,J$,  where $a_{[1]}=A$ and $a_{[J+1]}=L$. Making use of (\ref{eq:susc}), the prevalence of immune individuals of age $a$ is now well approximated by \citep{Anderson1991}:
\begin{equation}\label{eq:pidisc}
\pi(a) = 1 - \exp \left(-\sum_{k=1}^{j-1}\lambda_{k}(a_{[k+1]}-a_{[k]}) - \lambda_{j}(a-a_{[j]}) \right),
\end{equation}
if $a$ belongs to the $j{\mbox{th}}$ age interval. Note that we allow the prevalence of immune individuals to vary continuously with age and that we do not summarize the binary seroprevalence outcomes into a proportion per age class. Further, the force of infection for age class $i$ equals ($i=1,\ldots,J$):
\begin{equation}
\lambda_i = \frac{ND}{L} \sum_{j=1}^{J} \beta_{ij} \left[ \exp \left(-\sum_{k=1}^{j-1}\lambda_k(a_{[k+1]}-a_{[k]}) \right)-\exp\left(-\sum_{k=1}^{j}\lambda_k(a_{[k+1]}-a_{[k]})\right) \right] ,
\label{foid}
\end{equation}
where $\beta_{ij}$ denotes the per capita rate at which an individual of age class $j$ makes an effective contact with a person of age class $i$, per year. The transmission rates $\beta_{ij}$ make up a $J \times J$ matrix, the so-called WAIFW-matrix.

Once the WAIFW-matrix is estimated, following \cite{Diekmann1990} and \cite{Farrington2001}, the basic reproduction number $R_0$ can be calculated as the dominant eigenvalue of the $J\times J$ next generation matrix with elements ($i,j=1,\ldots,J$):
\begin{eqnarray} \label{eq:nextgen}
\frac{ND}{L} \left(a_{[i+1]}-a_{[i]} \right) \beta_{ij}.
\end{eqnarray}
$R_0$ represents the number of secondary cases produced by a typical infected person during his or her entire period of infectiousness, when introduced into an entirely susceptible population with the exception of newborns who are passively immune through maternal antibodies.
In the next section, we illustrate the traditional approach of imposing mixing patterns to estimate the WAIFW-matrix from seroprevalence data.

\subsection{Imposing mixing patterns} \label{sec:mixing}

The traditional approach of \cite{Anderson1991} imposes different, somewhat ad hoc, mixing patterns on the WAIFW-matrix. Note that, in the previous section, we ended up with a system of $J$ equations with $J\times J$ unknown parameters (\ref{foid}) and thus restrictions on these patterns are necessary. Among the proposals in the literature, one distinguishes between several mixing assumptions such as homogeneous mixing ($\beta(a,a^{\prime})=\beta$), proportional mixing ($\exists~u: \beta(a,a^{\prime})=u(a)u(a^{\prime})$), separable mixing ($\exists~u,v: \beta(a,a^{\prime})=u(a)v(a^{\prime})$) and symmetry ($\beta(a,a^{\prime})=\beta(a^{\prime},a)$). Note that the latter two mixing assumptions require additional restrictions to be made. As illustrated by \cite{Greenhalgh1994} and \cite{Van2009}, the structure imposed on the WAIFW-matrix has a high impact on the estimate of $R_0$.  In this section, we assume the transmission rates to be constant within six discrete age classes ($J=6$). We follow \cite{Anderson1991,Van2009,Ogunjimi2009} and consider the following mixing patterns, based on prior knowledge of social mixing behavior, to model the WAIFW-matrix for VZV:
\begin{eqnarray}
&&W_{1}=\left(\begin{array}{cccccc} \beta_1 & \beta_6 & \beta_6 & \beta_6 & \beta_6 & \beta_6 \\ \beta_6 & \beta_2 & \beta_6 & \beta_6 & \beta_6 & \beta_6 \\ \beta_6 & \beta_6 & \beta_3 & \beta_6 & \beta_6 & \beta_6 \\ \beta_6 & \beta_6 & \beta_6 & \beta_4 & \beta_6 & \beta_6 \\ \beta_6 & \beta_6 & \beta_6 & \beta_6 & \beta_5 & \beta_6 \\ \beta_6 & \beta_6 & \beta_6 & \beta_6 & \beta_6 & \beta_6 \\ \end{array}\right), ~~ W_{2}=\left(\begin{array}{cccccc} \beta_1 & \beta_1 & \beta_3 & \beta_4 & \beta_5 & \beta_6 \\ \beta_1 & \beta_2 & \beta_3 & \beta_4 & \beta_5 & \beta_6 \\ \beta_3 & \beta_3 & \beta_3 & \beta_4 & \beta_5 & \beta_6 \\ \beta_4 & \beta_4 & \beta_4 & \beta_4 & \beta_5 & \beta_6 \\ \beta_5 & \beta_5 & \beta_5 & \beta_5 & \beta_5 & \beta_6 \\ \beta_6 & \beta_6 & \beta_6 & \beta_6 & \beta_6 & \beta_6 \\ \end{array}\right) \nonumber \\
&&W_{3}=\left(\begin{array}{cccccc} \beta_1 & \beta_1 & \beta_1 & \beta_4 & \beta_5 & \beta_6 \\ \beta_1 & \beta_2 & \beta_3 & \beta_4 & \beta_5 & \beta_6 \\ \beta_1 & \beta_3 & \beta_3 & \beta_4 & \beta_5 & \beta_6 \\ \beta_4 & \beta_4 & \beta_4 & \beta_4 & \beta_5 & \beta_6 \\ \beta_5 & \beta_5 & \beta_5 & \beta_5 & \beta_5 & \beta_6 \\ \beta_6 & \beta_6 & \beta_6 & \beta_6 & \beta_6 & \beta_6 \\ \end{array}\right), ~~ W_{4}=\left(\begin{array}{cccccc} \beta_1 & \beta_1 & \beta_1 & \beta_1 & \beta_1 & \beta_1 \\ \beta_2 & \beta_2 & \beta_2 & \beta_2 & \beta_2 & \beta_2 \\ \beta_3 & \beta_3 & \beta_3 & \beta_3 & \beta_3 & \beta_3 \\ \beta_4 & \beta_4 & \beta_4 & \beta_4 & \beta_4 & \beta_4 \\ \beta_5 & \beta_5 & \beta_5 & \beta_5 & \beta_5 & \beta_5 \\ \beta_6 & \beta_6 & \beta_6 & \beta_6 & \beta_6 & \beta_6 \\ \end{array}\right) \label{mixing} \\
&&W_{5}=\left(\begin{array}{cccccc} \beta_1 & \beta_6 & \beta_6 & \beta_6 & \beta_6 & \beta_6 \\ \beta_6 & \beta_2 & \beta_6 & \beta_6 & \beta_6 & \beta_6 \\ \beta_6 & \beta_6 & \beta_3 & \beta_6 & \beta_6 & \beta_6 \\ \beta_6 & \beta_6 & \beta_6 & \beta_4 & \beta_6 & \beta_6 \\ \beta_6 & \beta_6 & \beta_6 & \beta_6 & \beta_5 & \beta_6 \\ \beta_6 & \beta_6 & \beta_6 & \beta_6 & \beta_6 & \beta_5 \\ \end{array}\right), ~~ W_{6}=\left(\begin{array}{cccccc} \beta_1 & 0 & 0 & 0 & 0 & 0 \\ 0 & \beta_2 & 0 & 0 & 0 & 0 \\ 0 & 0 & \beta_3 & 0 & 0 & 0 \\ 0 & 0 & 0 & \beta_4 & 0 & 0 \\ 0 & 0 & 0 & 0 & \beta_5 & 0 \\ 0 & 0 & 0 & 0 & 0 & \beta_6 \\ \end{array}\right).
\nonumber
\end{eqnarray}

In order to estimate the transmission parameters $\boldsymbol{\beta} = (\beta_1,\ldots,\beta_6)^{T}$ from seroprevalence data, we follow an iterative procedure from \cite{Farrington2001} and \cite{Kanaan2005}. First, one assumes plausible starting values for $\boldsymbol{\beta}$ and solves (\ref{foid}) iteratively for the piecewise constant force of infection $\boldsymbol{\lambda}= (\lambda_1,\ldots,\lambda_6)^{T}$, which in its turn can be contrasted to the serology. Second, this procedure is repeated under the constraint $\boldsymbol{\beta} \geq \boldsymbol{0}$, until the Bernoulli loglikelihood
\begin{equation*}
\sum_{i=1}^{n} \bigl\{y_i\log[\pi(a_i)]+(1-y_i)\log[1-\pi(a_i)]\bigr\},
\end{equation*}
has been maximized. Here, $n$ denotes the size of the serological data set, $y_i$ denotes a binary variable indicating whether subject $i$ had experienced infection before age $a_i$ and the prevalence $\pi(a_i)$ is obtained from (\ref{eq:pidisc}).

\subsection{Application to the data} \label{sec:vzvappl1}

For the remainder of the paper, the following parameters, specific for Belgium anno 2003 \citep{Eurostat2007,Federale2006}, are kept constant when estimating the WAIFW-matrix and $R_0$: size of the population aged 0 to 80 years, $N = 9943749$, and life expectancy at birth, $L=80$. The mean duration of infectiousness for VZV is taken $D=7/365$. Type I mortality and type I maternal antibodies with age $A=0.5$, are assumed. Removing individuals younger than 6 months, the size of the serological data set becomes $n=2649$.

\begin{figure}[ht]
\begin{center}
\includegraphics[width=6cm]{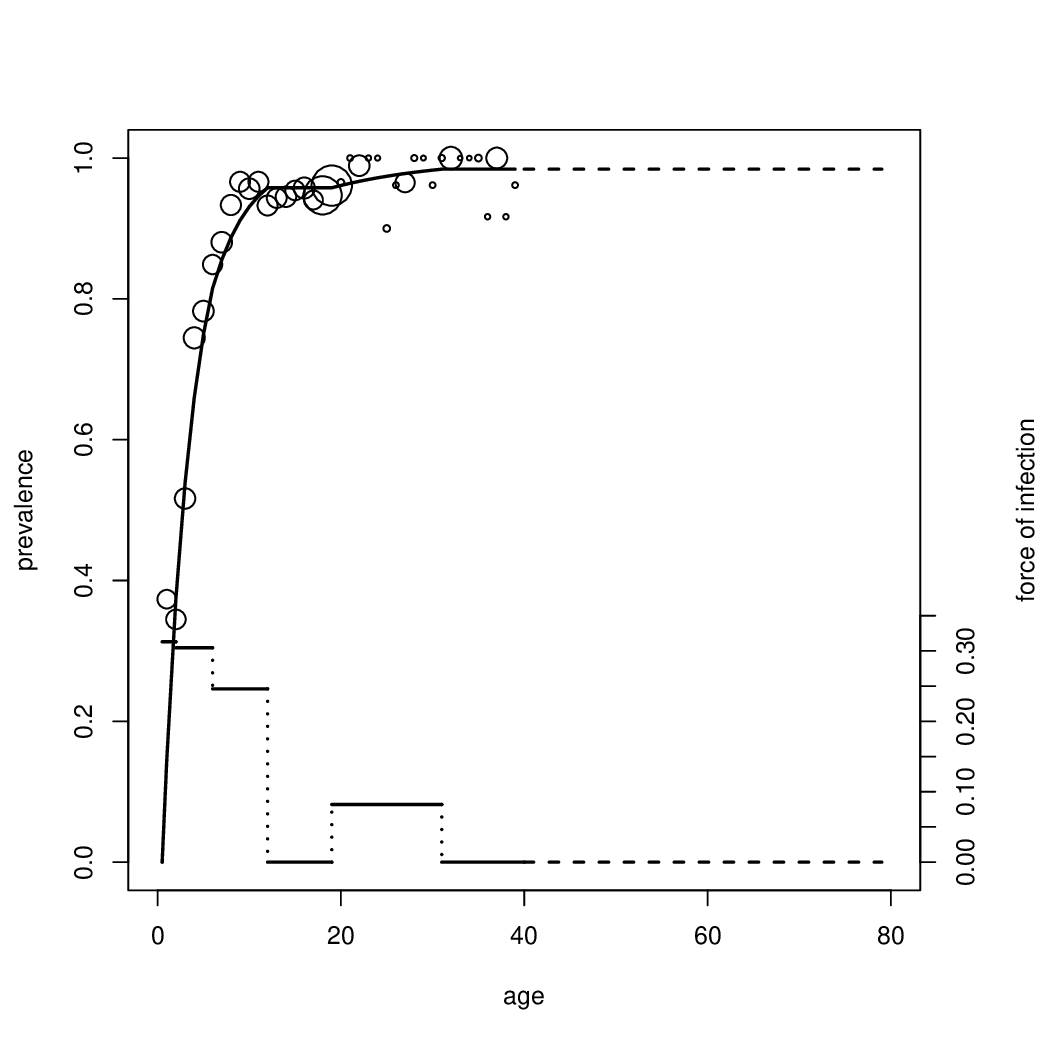}
\caption{Estimated prevalence (upper curve) and force of infection (lower curve) for VZV assuming a piecewise constant force of infection. The dots represent the observed serological data with size proportional to the corresponding sample size. The dashed lines are used to indicate the estimated prevalence and force of infection for the age interval $[40,80)$, which lacks serological information.} \label{fig:pcfit}
\end{center}
\end{figure}

In this application, the population is divided into six age classes taking into account the schooling system in Belgium, following \cite{Van2009}:  $(0.5,2)$, $[2,6)$, $[6,12)$, $[12,19)$, $[19,31)$, $[31,80)$. The last age class has a wide range because the serological data set only contains information for individuals up till 40 years. The following ML-estimate for $\boldsymbol{\lambda}$ is obtained assuming a piecewise constant force of infection and using constrained optimization to ensure monotonicity ($\pi'(a) \geq 0$): $\boldsymbol{\hat{\lambda}}^{\text{ML}}=(0.313, 0.304, 0.246, 0, 0.082, 0)^{T}$. A graphical display of the fit is presented in Figure~\ref{fig:pcfit} and a dashed line is used to indicate the estimated prevalence and force of infection for the age interval $[40,80)$ which lacks serological information.

During the estimation process, non-identifiability problems occur for mixing patterns $W_1$, $W_5$ and $W_6$, which is related to the fact that $\hat{\lambda}_4^{\text{ML}}=\hat{\lambda}_6^{\text{ML}}=0$. Therefore, these mixing patterns are left from further consideration. For the remaining three, ML-estimates for $\boldsymbol{\beta}$ and $R_0$ are presented in Table~\ref{table:WAIFW}. Note that mixing pattern $W_4$ has a regular configuration for the data, whereas $W_2$ and $W_3$ are non-regular since unconstrained ML-estimation induces negative estimates for $\beta_4$ \citep{Farrington2001}. The estimate of $R_0$ ranges from 3.37 to 4.21. A 95\% bootstrap-based percentile confidence interval for $R_0$ is presented as well, applying a non-parametric bootstrap by taking $B=1000$ samples with replacement from the serological data. The fit of the three mixing patterns can be compared using model selection criteria, such as AIC and BIC \citep{Schwarz1978}. As can be seen from Table~\ref{table:WAIFW}, the AIC-values (equivalent to BIC here) are virtually equal and do not provide any basis to guide the choice of a mixing pattern.

\begin{table}
\caption{\label{table:WAIFW} Estimates for the transmission parameters (multiplied by $10^{4}$) and for $R_0$, obtained by imposing mixing patterns $W_2$, $W_3$ and $W_4$ on the WAIFW-matrix.}
\centering
\fbox{%
\begin{tabular}{cccccccccc}
 & $\hat{\beta}_1$ & $\hat{\beta}_2$ & $\hat{\beta}_3$ & $\hat{\beta}_4$ & $\hat{\beta}_5$ & $\hat{\beta}_6$ & $\widehat{R}_0$ & 95\% CI for $R_0$ & AIC \\
\hline \vspace{-0.3cm} \\
$W_2$ & 1.413 & 1.335 & 1.064 & 0.000  & 0.343  & 0.000  & 3.51 & [3.07, 13.42] & 1372.819\\
$W_3$ & 1.362 & 1.441 & 0.873 & 0.000 & 0.343 & 0.000 & 3.37 & [2.81, 13.38] & 1372.819 \\
$W_4$ & 1.334 & 1.298 & 1.049 & 0.000 & 0.349 & 0.000 & 4.21 & [3.69, 13.13] & 1372.756\\
\end{tabular}}
\end{table}

Note that these results differ somewhat from those obtained by \cite{Van2009}, where a different data set for VZV serology was used, collected from a large laboratory in the city of Antwerp between October 1999 and April 2000.

\section{Estimation of $R_0$ using data on social contacts}

\subsection{Constant proportionality of the transmission rates}

In the previous section, we have illustrated some caveats involved in the traditional approach of imposing mixing patterns on the WAIFW-matrix. In general, the choice of the structures as well as the choice of the age classes are somewhat ad hoc. Since evidence for mixing patterns is thought to be found in social contact data, i.e.~governing contacts with high transmission potential, an alternative approach to estimate transmission parameters has emerged: augmenting seroprevalence data with data on social contacts.  In \cite{Wallinga2006}, it was argued that $\beta(a,a')$ is proportional to $c(a,a')$, the per capita rate at which an individual of age $a'$ makes contact with a person of age $a$, per year:
\begin{equation}\label{eq:qc}
\beta(a,a^{\prime})=q \cdot c(a,a^{\prime}).
\end{equation}
We will refer to this assumption as the `constant proportionality' assumption, since $q$ represents a constant disease-specific factor.
Translating this assumption into the discrete framework with age classes  $(a_{[1]},a_{[2]}), [a_{[2]},a_{[3]}), \ldots, [a_{[J]},a_{[J+1]})$,  is straightforward ($i,j=1,\ldots,J$): $\beta_{ij}=q \cdot c_{ij}$,
where $c_{ij}$ denotes the per capita rate at which an individual of age class $j$ makes contact with a person of age class $i$, per year.

The proportionality factor and the contact rates are not identifiable from serological data only. Therefore, in order to estimate the WAIFW-matrix, one first needs to estimate the contact rates $c_{ij}$ using social contact data. Following the Belgian contact survey, `making contact with' is then defined as a two-way conversation of at least three words in each others proximity and/or any sort of physical skin-to-skin touching (Section \ref{contact data}). In Section \ref{sec:refine}, we will refine on this definition and consider specific types of contact with high transmission potential.  In a second step, keeping the estimated contact rates fixed, we estimate the proportionality factor from serological data using the estimation method described in Section \ref{sec:mixing}.

\subsection{Estimating contact and transmission rates}

Consider the random variable $Y_{ij}$, i.e. the number of contacts in age class $j$ during one day as reported by a respondent in age class $i$ ($i,j=1,\ldots,J$), which has observed values $y_{ij,t}$, $t=1,\ldots,T_i$, where $T_i$ denotes the number of participants in the contact survey belonging to age class $i$. Now define $m_{ij} = E(Y_{ij})$, i.e. the mean number of contacts in age class $j$ during one day as reported by a respondent in age class $i$. The elements $m_{ij}$ make up a $J \times J$ matrix, which is called the `social contact matrix'. Now, the contact rates $c_{ij}$ are related to the social contact matrix as follows:
\[ c_{ij} = 365 \cdot \frac{m_{ji}}{w_i}, \]
where $w_i$ denotes the population size in age class $i$, obtained from demographical data. When estimating the social contact matrix, the reciprocal nature of contacts needs to be taken into account \citep{Wallinga2006}:
\begin{equation}
m_{ij}w_i=m_{ji}w_j,
\label{eq:recipr}
\end{equation}
which means that the total number of contacts from age class $i$ to age class $j$ must equal the total number of contacts from age class $j$ to age class $i$.

\subsubsection{Bivariate smoothing}
The elements $m_{ij}$ of the social contact matrix are estimated from the contact data using a bivariate smoothing approach as described by \cite{Wood2006}. In contrast with the maximum likelihood approach as presented by \cite{Wallinga2006}, the average number of contacts is modeled as a two-dimensional continuous function over age of respondent and contact, giving rise to a `contact surface'. The basis is a tensor-product spline derived from two smooth functions of the respondent's and contact's age, ensuring flexibility:
\begin{eqnarray}  \label{eq:smooth}
Y_{ij} \sim \mbox{NegBin}(m_{ij},k), \mbox{ where } g(m_{ij}) = \sum_{\ell=1}^{K} \sum_{p=1}^{K} \delta_{\ell p} b_{\ell}(a_{[i]}) d_p(a_{[j]}),
\end{eqnarray}
where $g$ is some link function, $\delta_{\ell p}$ are unknown parameters, and $b_{\ell}$ and $d_p$ are known basis functions for the marginal smoothers.
To allow for overdispersion, we assume that the contact counts $Y_{ij}$ are independently negative binomial distributed with mean $m_{ij}$, dispersion parameter $k$ and variance $m_{ij} + m_{ij}^2/k$.

The basis dimension, $K$, should be chosen large enough in order to fit the data well, but small enough to maintain reasonable computational efficiency \citep{Wood2006}. For tensor-product smoothers, the upper limit of the degrees of freedom is given by the product of the $K$ values provided for each marginal smooth, minus one, for the identifiability constraint. However, the actual effective degrees of freedom are also controlled by the degree of penalization selected during fitting.

Thin plate regression splines are used to avoid the selection of knots and a log link is used in model (\ref{eq:smooth}). Diary weights, as discussed in Section~\ref{contact data}, are taken into account in the smoothing process. By applying a smooth-then-constrain-approach as proposed by \cite{Mammen2001}, the reciprocal nature of contacts (\ref{eq:recipr}) is taken into account.

\subsubsection{Estimating the contact rates} \label{sec:smooth}
\begin{figure}[ht]
\centering
\includegraphics[width=6cm]{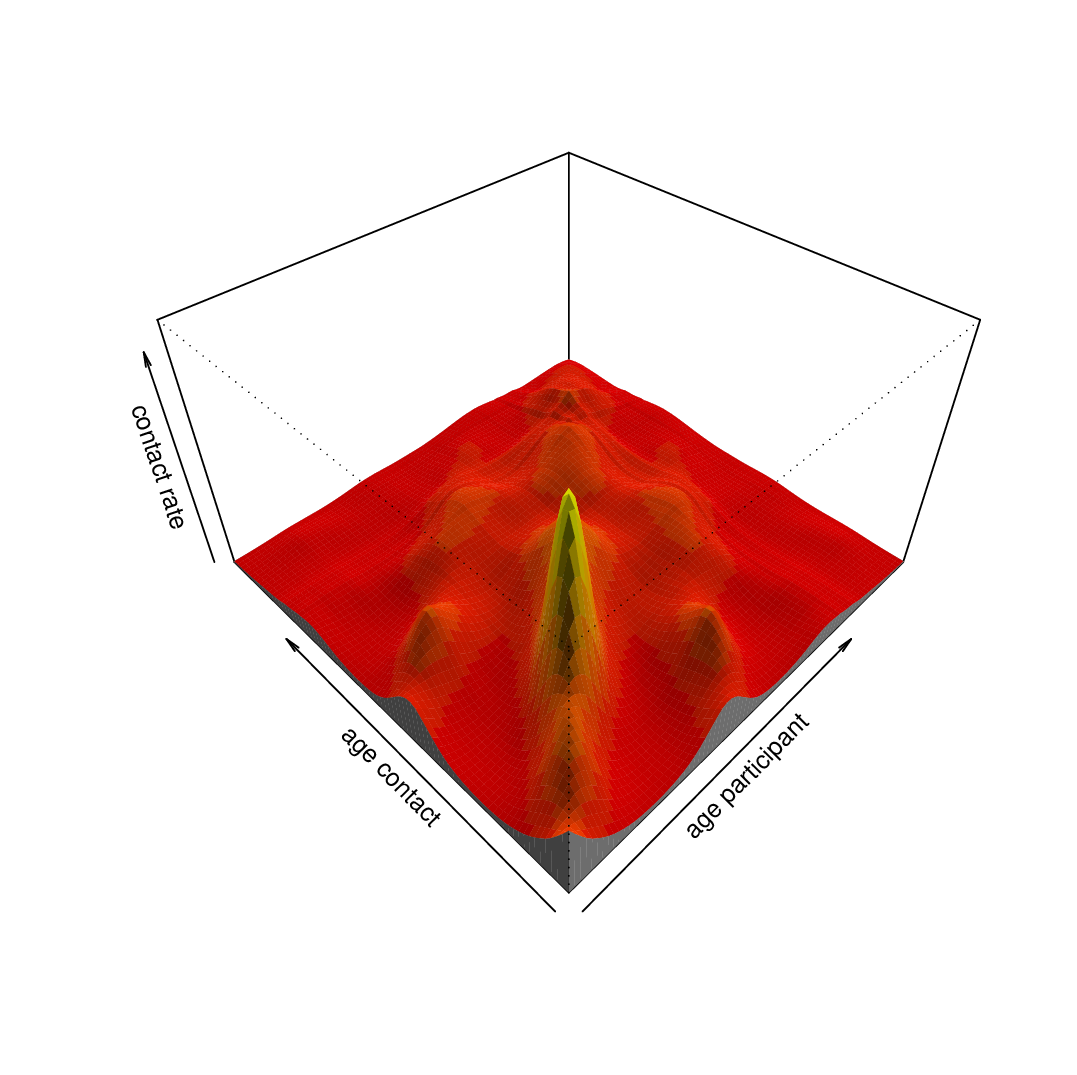} \hspace{1cm}
\includegraphics[width=6cm]{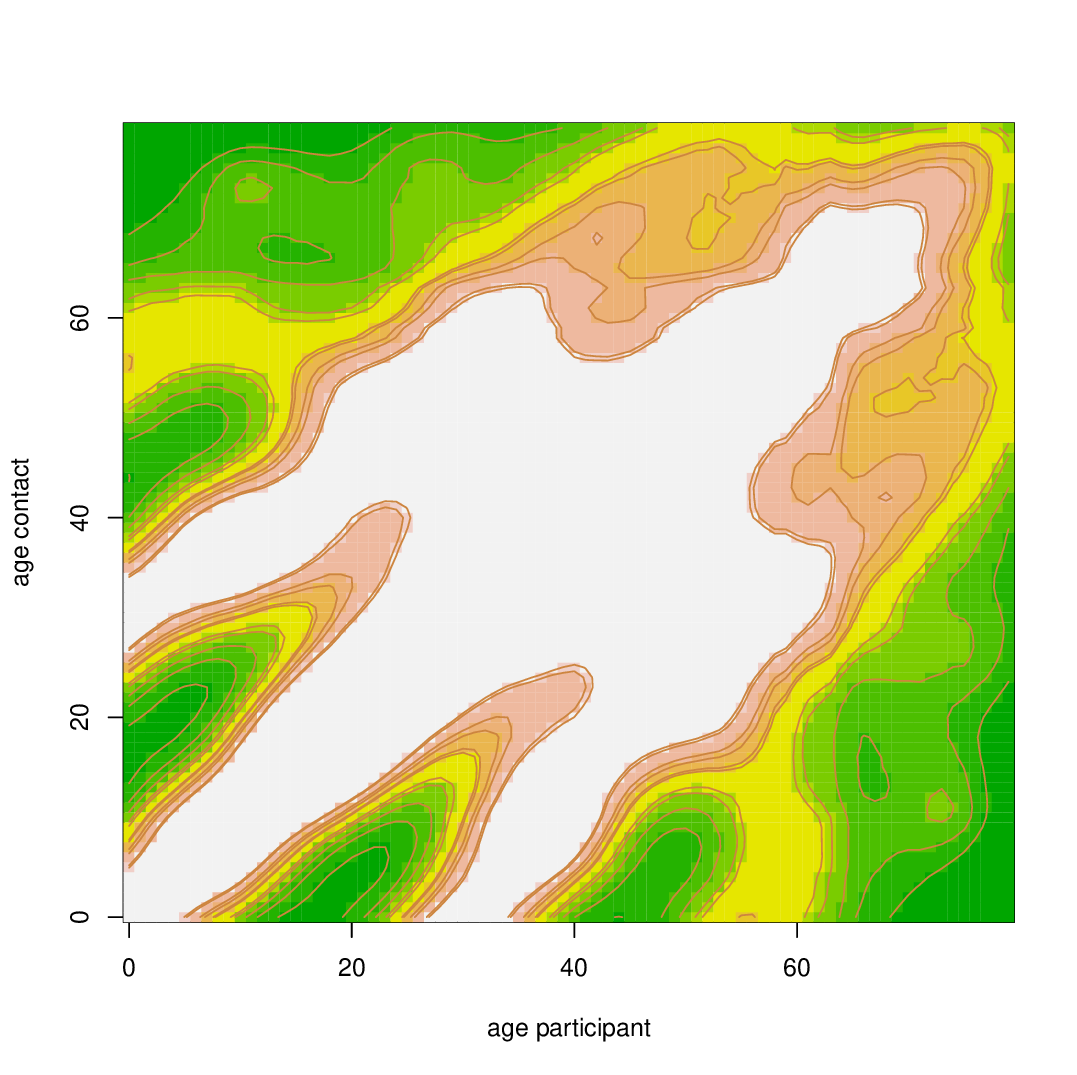}
\caption{Perspective (left) and image (right) plot of the estimated contact rates $c_{ij}$ obtained with bivariate smoothing. The $X$- and $Y$-axis represent age of the respondent and age of the contact, respectively.} \label{fig:fitcont}
\end{figure}

The smoothing is performed in R with the \verb+gam+ function from the \verb+mgcv+ package \citep{Wood2006}, considering one year age intervals, $[0,1),[1,2),\ldots,[100,101)$. An informal check (by comparing the estimated degrees of freedom and the basis dimension) shows that $K=11$ is a satisfactory basis dimension choice for the Belgian contact data. In Figure~\ref{fig:fitcont}, the estimated contact surface obtained with the bivariate smoothing approach, is displayed. The smoothing approach seems well able to capture important features of human contacting behavior.
Three components clearly arise in the smoothed contact surface. First of all, one can see a pronounced assortative structure on the diagonal, representing high contact rates between individuals of the same age. Second, an off-diagonal parent-child component comes forward, reflecting a very natural form of contact between parents and children, which might be important in modeling certain childhood infections such as parvovirus B19 \citep{Mossong2008a}. Finally, there even seems to be evidence for a grandparent-grandchild component.

Except for the assortativeness, these features are not reflected by the contact rates, estimated by maximizing the likelihood of the `saturated model' proposed by \cite{Wallinga2006}, considering the same six age classes used in Section \ref{sec:vzvappl1} (results omitted here). Furthermore, AIC and BIC criteria indicate the smoothing method to outperform \cite{Wallinga2006}'s saturated model, showing improved estimation of the contact surface using nonparametric techniques.

\subsubsection{Estimating $R_0$} \label{sec:vzvappl2}
\begin{figure}[!ht]
\centering
\includegraphics[width=6cm]{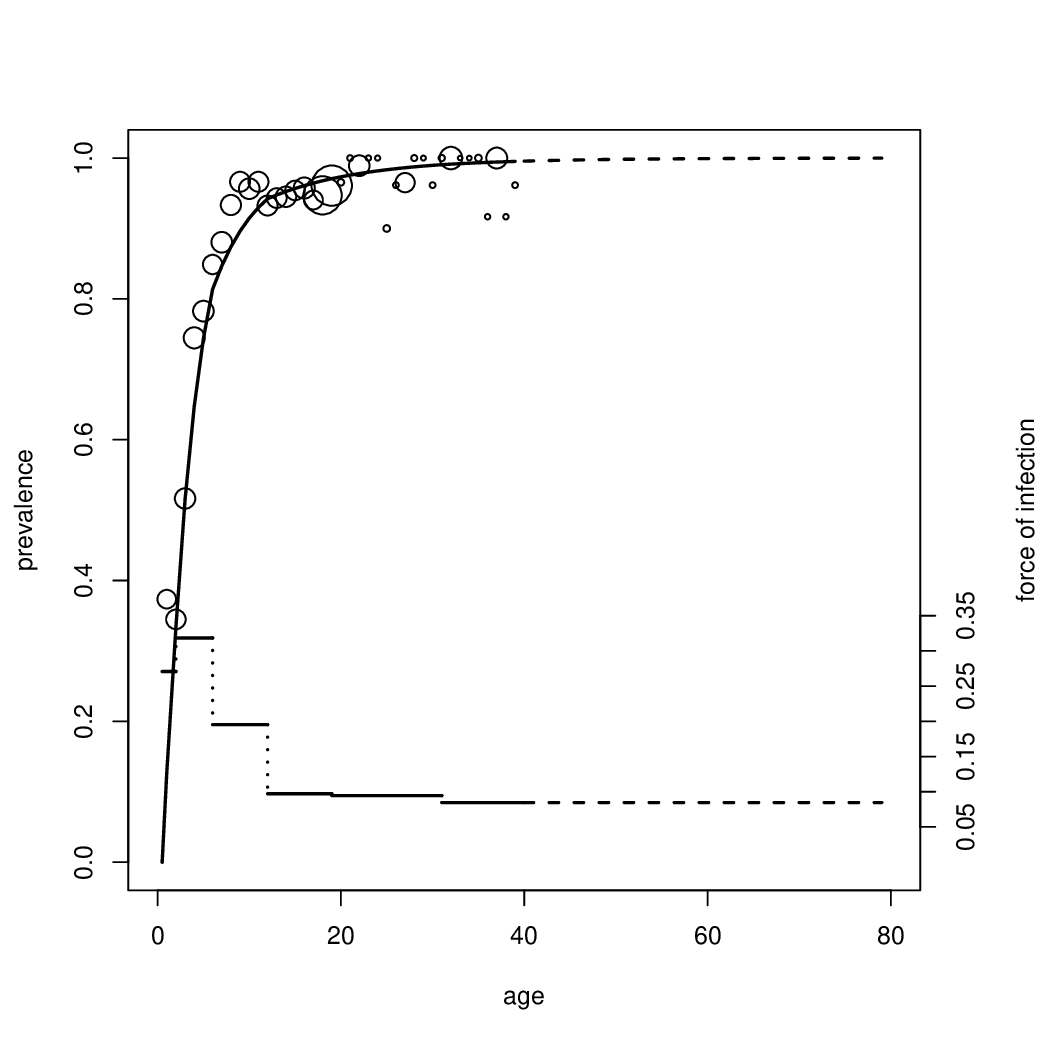} \hspace{1cm}
\includegraphics[width=6cm]{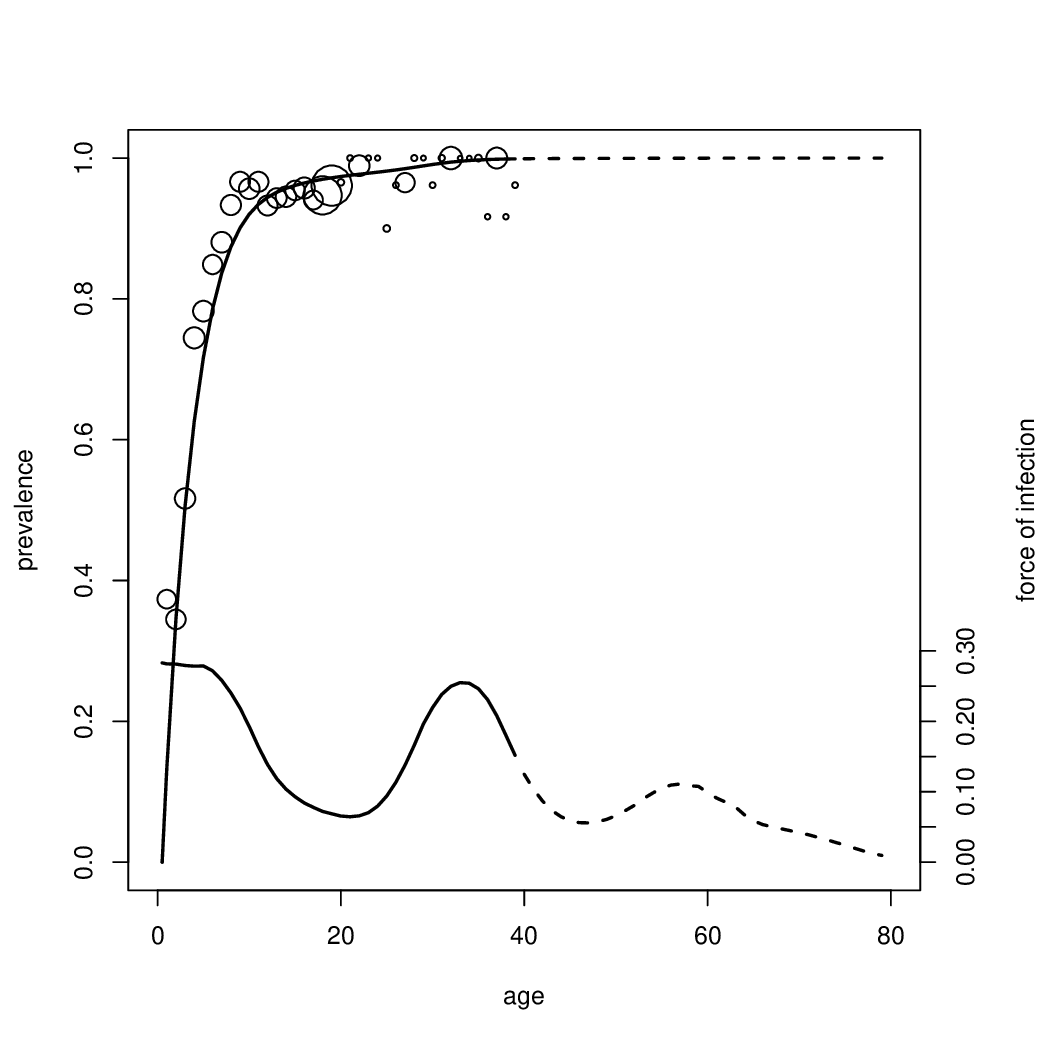}
\caption{Estimated prevalence (upper curve) and force of infection (lower curve) obtained from contact rates estimated using maximum likelihood for \cite{Wallinga2006}'s saturated model (left) and using bivariate smoothing (right).} \label{fig:fitprev}
\end{figure}

Under the constant proportionality assumption (\ref{eq:qc}), we are now able to estimate the WAIFW-matrix for VZV using serological data.  Keeping the estimated contact rates $\hat{c}_{ij}$ fixed, we estimate the proportionality factor $q$ using the estimation method described in Section~\ref{sec:mixing}. In Table~\ref{table:q1}, estimates for $q$ and $R_0$ together with their corresponding 95\%~profile likelihood confidence intervals, and AIC-values, are presented for the bivariate smoothing approach and the `saturated model' proposed by \cite{Wallinga2006}. The results are fairly similar, though the saturated model induces a smaller AIC-value compared to the smoothing approach. As can be seen from both model fits in Figure~\ref{fig:fitprev}, contact rate estimates between children will mainly determine the fit to the serological data, limiting the advantage of a better contact surface estimate. Note that the 95\%~confidence intervals in Table~\ref{table:q1} are implausibly narrow, resulting from the fact that the estimated contact rates are held constant.

\begin{table}
\caption{\label{table:q1} ML-estimates for the proportionality factor and $R_0$, obtained from contact rates estimated by bivariate smoothing and \cite{Wallinga2006}'s saturated model, assuming constant proportionality.}
\centering \fbox{%
\begin{tabular}{cccrcc}
Model for $c_{ij}$ & $\hat{q}$ & 95\% CI for $q$ & $\widehat{R}_0$ & 95\% CI for $R_0$ & AIC \\
\hline \vspace{-0.3cm} \\
Smoothing & 0.132 & [0.124, 0.140] & 15.69 & [14.74, 16.69] & 1386.618 \\
Saturated & 0.124  &  [0.117, 0.132] & 14.08  & [13.26, 14.94] & 1377.146 \\
\end{tabular}}
\end{table}

\subsection{Refinements to the social contact data approach}

The aim is to clearly disentangle the WAIFW-matrix into the contact process and the transmission potential. Therefore, in the following, contact rates are estimated using a bivariate smoothing approach, since this method outperforms the saturated model estimated using maximum likelihood as proposed by \cite{Wallinga2006} (Section~\ref{sec:smooth}). Following \cite{Ogunjimi2009} and \cite{Melegaro2009}, contacts with high transmission potential are filtered from the social contact data. Further, to improve statistical inference, we present a non-parametric bootstrap approach, explicitly accounting for all sources of variability.

\subsubsection{Contacts with high transmission potential} \label{sec:refine}

The aim is to trace the type of contact which is most likely to be responsible for VZV transmission, hereby exploiting the following details provided on each contact: duration and type of contact, which is either close or non-close (Section~\ref{contact data}). Five types of contact are considered and we will explore which one induces the best fit to the serological data. First, the contact rates $c(a,a')$ are estimated using the complete contact data set as we did in Section \ref{sec:vzvappl2} and further, four specific types of contact with high transmission potential for VZV are selected according to \cite{Ogunjimi2009} and \cite{Melegaro2009}:
\begin{center}
\begin{tabular}{ccl}
Model & Parameter & Type of contact \\
\hline
$C_1$ & $q_1$ & all contacts \\
$C_2$ & $q_2$ & close contacts \\
$C_3$ & $q_3$ & close contacts $> 15$ minutes	\\
$C_4$ & $q_4$ & close contacts and non-close contacts $> 1$ hour\\
$C_5$ & $q_5$ & close contacts $> 15$ minutes and non-close contacts $>1$ hour \\
\end{tabular}
\end{center}

Assuming constant proportionality, maximum likelihood estimates for the transmission parameters $q_k$ ($k=1,\ldots,5$) and for the basic reproduction number $R_0$ together with their corresponding 95\% profile likelihood confidence intervals (first entry), are presented in Table~\ref{tab:q}. For each model $C_k$, the AIC-value, AIC difference $\Delta_k = \mbox{AIC}_k - \mbox{AIC}_{\min}$, Akaike weight
\[w_k = \frac{ \exp(-\frac{1}{2} \Delta_k)}{\sum \limits_{\ell} \exp(-\frac{1}{2} \Delta_{\ell})},\]
and evidence ratio (ER) $w_{\min}/w_k$, are calculated following \cite{Burnham2002}, where $\mbox{AIC}_{\min}$ and $w_{\min}$ correspond to the model with the smallest AIC value. Recall that the AIC is an estimate of the expected, relative Kullback-Leibler (K-L) distance, whereas the K-L distance embodies the information lost when an approximating model is used instead of the unknown, true model. A given Akaike weight $w_k$ is considered as the weight of evidence in favor of a model $k$ being the actual K-L best model for the situation at hand, given the data and the set of candidate models considered.

According to the AIC-criterion, although AIC differences are minor, the contact matrix consisting of close contacts longer than 15 minutes (model $C_3$) implies the best fit to the seroprevalence data. A graphical representation of the estimated prevalence and force of infection is omitted here, since the result is very close to the one obtained for model $C_1$ in Figure~\ref{fig:fitprev}. Further, there is evidence for model $C_5$ as well, having an Akaike weight of 0.329 and an evidence ratio of 1.7. The latter model adds non-close contacts longer than one hour to model $C_3$ and therefore these models are closely related.

\begin{table}
\caption{\label{tab:q}ML-estimates for the proportionality factor and $R_0$,  95\% profile likelihood confidence intervals (first entry), 95\% bootstrap-based percentile confidence intervals (second entry) and several measures related to model selection,  obtained from contact rates estimated using bivariate smoothing, considering different types of contact $C_1$-$C_5$, assuming constant proportionality. }
\centering\fbox{%
\small
\begin{tabular}{cccrccrcr}
Model & $\hat{q}_k$ & 95\% CI for $q_k$ &	$\widehat{R}_0$ & 95\% CI for $R_0$ & AIC & $\Delta_k$ & $w_k$ & ER \\
\hline \vspace{-0.2cm} \\
$C_1$ & 0.132 & [0.124, 0.140]  & 15.69	& [14.74, 16.69]& 1386.618 & 11.660 & 0.002& 340.4 \\
 &  & [0.103, 0.175]  & & [12.34, 21.41] & &  & &  \\
$C_2$ &	0.160 & [0.150, 0.169]	& 10.24	& [9.65, 10.85]	& 1379.581 & 4.623 & 0.057& 10.1\\
&  & [0.126, 0.208]  & 	&  [8.21, 13.68]& &  & &  \\
$C_3$ & 0.173 & [0.163, 0.184]	& 8.68	& [8.18, 9.20]	& 1374.958 & 0.000 & 0.574& 1.0\\
&  & [0.133, 0.221]  & 	& [6.89, 11.34] & &  & &  \\
$C_4$ & 0.145 & [0.136, 0.154]	& 11.73	& [11.05, 12.47]& 1380.354 & 5.396 & 0.039& 14.9 \\
&  & [0.113, 0.188]  & 	& [9.41, 15.95]& &  & &  \\
$C_5$ & 0.156 & [0.147, 0.166]	& 10.40	& [9.79, 11.04] & 1376.068 & 1.110 & 0.329& 1.7\\
&  & [0.119, 0.204]  & 	&[8.05, 14.10]& &  & &  \\
\end{tabular}}
\end{table}
\normalsize

\subsubsection{Non-parametric bootstrap} \label{sec:bootstrap}

We explicitly acknowledge that up till now, by keeping the estimated contact rates fixed, we have ignored the variability originating from the contact data. In order to assess sampling variability for the social contact data and the serological data altogether, we will use a non-parametric bootstrap approach. Furthermore, building in a randomization process, uncertainty concerning age is accounted for. After all, in the social contact data, ages of respondents are rounded up, which is also the case for some individuals in the serological data set. Concerning the age of contacts, a lower and upper age limit is given by the respondents. Instead of using the mean value of these age limits, a random draw is now taken from the uniform distribution on the corresponding age interval. In summary, each bootstrap cycle consists of the following six steps:
\begin{enumerate}
\item randomize ages in the social contact data and the serological data set;
\item take a sample with replacement from the respondents in the social contact data;
\item recalculate diary weights based on age and household size of the selected respondents;
\item estimate the social contact matrix (smooth-then-constrain approach);
\item take a sample with replacement from the serological data;
\item estimate the transmission parameters and $R_0$.
\end{enumerate}
This bootstrap approach allows one to calculate bootstrap confidence intervals for the transmission parameters and for the basic reproduction number, which take into account all sources of variability.

The impact on statistical inference is now illustrated for the models considered in the previous section. Nine hundred bootstrap samples are taken from the contact data and from the serological data simultaneously, while ages are being randomized. Merely $B=587$ bootstrap samples lead to convergence in all five smoothing procedures, which might be induced by the sparse structure of the contact data.  However, by individual monitoring of non-converging \verb+gam+ functions, convergence was reached after all and a comparison of the bootstrap results showed little difference whether or not these samples were included. 95\% percentile confidence intervals for $q$ and $R_0$ are calculated based on the $B=587$ bootstrap samples (see Table~\ref{tab:q}, second entry). Taking into account sampling variability for the social contact data has a noticeable impact, as can be seen from the wider 95\% confidence intervals.

\section{Age-dependent proportionality of the transmission rates} \label{sec:agedep}

The proportionality factor $q$ might depend on several characteristics related to susceptibility and infectiousness, which could be ethnic-, climate-, disease- or age-specific. Examples of age-specific characteristics related to susceptibility and infectiousness include the mean infectious period, mucus secretion and hygiene. In the situation of seasonal and pandemic influenza this has been established and used in realistic simulation models (see e.g. \cite{Cauchemez2004} and \cite{Longini2005}). Furthermore, the conversational and physical contacts reported in the diaries serve as proxies of those events by which an infection can be transmitted. For example, sitting close to someone in a bus without actually touching each other, may also lead to transmission of infection. In light of these discrepancies, $q$ can be considered as an age-specific adjustment factor which relates the true contact rates underlying infectious disease transmission to the social contact proxies.

In view of this, we will explore whether $q$ varies with age, an assumption we will refer to as `age-dependent proportionality':
\begin{equation}\label{eq:qac}
\beta(a,a^{\prime})=q(a,a^{\prime})\cdot c(a,a^{\prime}),
\end{equation}
which in the discrete framework turns into: $\beta_{ij}=q_{ij}\cdot c_{ij}$ ($i,j=1,\ldots,J$). In the previous section, it was observed that, under the constant proportionality assumption, close contacts longer than 15 minutes imply the best fit to the serological data for VZV. Therefore in the following, the contact rate is modeled using close contacts longer than 15 minutes and we will elaborate on this particular model by assuming age dependence. First, discrete structures are applied in order to model $q$ as an age-dependent proportionality factor and second, `continuous' loglinear regression models are considered for the same purpose. Finally, we assess the level of model selection uncertainty and calculate a model averaged estimate for the basic reproduction number.

\subsection{Discrete structures} \label{sec:discrete}

The proportionality factor $q_{ij}$ is now allowed to differ between age classes. Discrete matrix structures, involving two transmission parameters $\gamma_1$ and $\gamma_2$, are explored in modeling $q_{ij}$. Five models are considered, which fit the following structures for $q_{ij}$ to the seroprevalence data:
\begin{eqnarray*}
&&M_1 = \left(\begin{array}{cc} \gamma_1 & \gamma_2 \\ \gamma_2 & \gamma_2 \end{array}\right), ~M_2=\left(\begin{array}{cc} \gamma_1 & \gamma_1 \\ \gamma_2 &\gamma_2 \end{array}\right), ~M_3=\left(\begin{array}{cc} \gamma_1 & \gamma_2 \\ \gamma_2 &\gamma_1 \end{array}\right), \\ &&M_4 =\left(\begin{array}{cc} \gamma_1 & 0 \\ 0 &\gamma_2 \end{array}\right), ~M_5=\left(\begin{array}{cc} \gamma_1 & \gamma_2 \\ \gamma_1 &\gamma_2 \end{array}\right).
\end{eqnarray*}
The population is divided into two age classes, namely $[0.5,12)$ and $[12,80)$, a choice based on the dichotomy of the population according to the schooling system in Belgium (Section~\ref{sec:vzvappl1}), yielding the smallest AIC-value. Note that higher order extensions, considering more parameters and/or number of age classes, were fitted to the serological data as well. The improvement in loglikelihood, however, does not outweigh the increase in the number of transmission parameters.

Notice that the structures of $M_1$-$M_5$ resemble the mixing patterns imposed on the WAIFW-matrix in the traditional \cite{Anderson1991} approach. We would like to emphasize that the method proposed here differs greatly from the latter, since the WAIFW-matrix is now estimated using the estimated contact rates: $\beta_{ij}=q_{ij}\cdot\hat{c}_{ij}$. Hence, in contrast with the approach of \cite{Anderson1991} who estimate $\beta_{ij}$ by fixing the structure of the mixing pattern, in our approach we estimate the contact pattern from the survey data and use several proportionality structures to select the best model from which the $\beta_{ij}$ are estimated.

Table~\ref{tab:candidate} displays ML-estimates for $\gamma_1$, $\gamma_2$ and the basic reproduction number $R_0$, together with their corresponding 95\% percentile confidence intervals ($B=603$ bootstrap samples converged out of 700).  For model $M_4$, $\gamma_2$ is non-identifiable, and unconstrained optimization of model $M_5$ would not lead to convergence.  According to the AIC-criterion, the remaining models fit equally well and are informative with respect to VZV transmission dynamics. Most likely, this is due to the fact that the main transmission routes for VZV are between children and from infectious children to susceptible adults, embodied by the first column $(\gamma_1,\gamma_2)^{T}$. The three models result in approximately the same estimates for $\gamma_1$ and $\gamma_2$ and consequently the differences in AIC are only minor.

It is clear from Table~\ref{tab:candidate} that we estimate a difference in transmissibility between those younger and older than 12 years (about 0.18 and 0.07, respectively). This difference cannot be solely explained by the estimated contact rates. A possible explanation is that when infectious children make close contact with susceptible children during a sufficient amount of time, the probability of effective VZV transmission is higher compared to the same situation with susceptible adults. Another potential cause is underreporting of contacts between children. After all, up to the age of eight, the contact diaries were filled in by the parents, which may have induced some reporting bias \citep{Hens2009a}.

\begin{table}
\caption{\label{tab:candidate}Candidate models for the proportionality factor together with ML-estimates for the transmission parameters and $R_0$,  95\% bootstrap-based percentile confidence intervals,  and several measures related to model selection.}
\centering\fbox{%
\small
\begin{tabular}{ccrcccccccr}
Model &  \multicolumn{2}{c}{Parameter} & 95\% CI & $\widehat{R}_0$ & 95\% CI for $R_0$ & $K$ & AIC & $\Delta_k$ & $w_k$ & ER \\
\hline \vspace{-0.2cm} \\
$C_3$ & $\hat{q}$ & 0.173 & [0.133, 0.221] &  8.68 & [6.89, 11.34] &  1 & 1374.958 &  8.884&  0.003 & 84.9 \\
$M_1$ & $\hat{\gamma}_1$ &  0.185 & [0.136, 0.244] &  4.79 & [4.15, 9.98] & 2 & 1366.306 &  0.232&  0.261 & 1.1 \\
& $\hat{\gamma}_2$ & 0.079 & [0.006, 0.196] & & & & & & & \\
$M_2$ & $\hat{\gamma}_1$ & 0.183 & [0.138, 0.240] & 5.37 & [4.47, 9.68] & 2 & 1366.285 &  0.211&  0.264 & 1.1\\
& $\hat{\gamma}_2$ & 0.078 & [0.006, 0.187] & & & & & & &\\
$M_3$ & $\hat{\gamma}_1$ & 0.185 & [0.136,  0.244] & 8.26 & [6.82, 11.25] & 2 & 1366.074 &  0.000&  0.293 & 1.0\\
& $\hat{\gamma}_2$ &  0.069 & [0.006, 0.199] & & & & & & &\\
$M_6$ & $\hat{\gamma}_0$ &-1.622 &  [-2.028, -1.212] & 5.79& [4.63, 12.60] & 2 & 1368.709 &  2.635&  0.079 & 3.7\\
& $\hat{\gamma}_1$ &  -0.023 & [-0.067, 0.016] & & & & & &  &\\
$M_7$ &  $\hat{\gamma}_0$ &  -1.720 &  [-2.441, -1.182] & 5.03 & [4.20, 1318.68] &  3 &  1368.325 &  2.251&  0.095 & 3.1 \\
& $\hat{\gamma}_1$ & 0.014 &  [-0.086, 0.305] & & & & &  & &\\
&  $\hat{\gamma}_2$ & -0.002 & [-0.024, 0.001] & & & & & & &\\
$M_8$ &  $\hat{\gamma}_0$ & -1.517 &  [-2.224, -0.446] & 3.55 & [1.76, 159.96] &  2 & 1374.324 &  8.250&  0.005 & 61.9 \\
& $\hat{\gamma}_1$ & -0.065 &  [-0.403, 0.064] & & & & & &\\
\end{tabular}}
\end{table}
\normalsize

\subsection{Continuous modeling} \label{sec:continuous}

As opposed to the previous, the proportionality factor $q(a,a')$ is now allowed to vary continuously over age. Loglinear regression models are considered for $q(a,a')$, since we expect an exponential decline of $q$ over $a$ due to hygienic habits as well as an exponential decline of $q$ over $a'$ due to decreasing mucus secretion. The following loglinear models are fitted to the data:
\begin{eqnarray*}
M_6:&  \log\{q(a)\} = &\gamma_0 + \gamma_1 a, \\
M_7:&  \log\{q(a)\} = &\gamma_0 + \gamma_1 a + \gamma_2 a^{2},\\
M_8:&  \log\{q(a')\} = &\gamma_0 + \gamma_1 a',	\\
M_9:&  \log\{q(a')\} = &\gamma_0 + \gamma_1 a' + \gamma_2 (a')^{2},\\
M_{10}:&  ~~\log\{q(a,a')\} = &\gamma_0 + \gamma_1 a + \gamma_2 a'.
\end{eqnarray*}
Model $M_6$ models $q$ as a first degree function of age of the susceptible and model $M_7$ allows for an additional quadratic effect of age, $a^{2}$. Models $M_8$ and $M_9$ are the analogue of $M_6$ and $M_7$ for age of the infectious person, $a'$. Finally, $M_{10}$ models $q$ as an exponential function of $a$ and $a'$ simultaneously. For model $M_9$, no convergence was obtained and model $M_{10}$ gives rise to an estimated proportionality factor which is exponentially increasing over $a'$, inducing unrealistically large estimates for $q$ at older ages.

Maximum likelihood estimates for the model parameters and the basic reproduction number $R_0$ are presented in Table~\ref{tab:candidate}, together with the corresponding 95\% percentile confidence intervals ($B=603$ bootstrap samples converged out of 700). According to the AIC-criterion, $M_6$ and $M_7$ fit equally well. Allowing the proportionality factor to vary by age of infectious persons, does not seem to substantially improve model fit, as can be seen by comparing the AIC-values of $C_3$ and $M_8$.

Clearly for models $M_7$ and $M_8$, the upper limits of the confidence intervals for $R_0$ are very large, as a consequence of estimated proportionality factors which are exponentially increasing over $a$ and $a'$, respectively. This result originates from two things: first, there is lack of serological information for individuals aged 40 and older, and second, VZV is highly prevalent in the population and most individuals become infected with VZV before the age of ten. Mathematically the latter means that from a certain age on, $\pi(a)\approx1$ and $\pi'(a)\approx0$, leading to an indeterminate force of infection $\lambda(a) = \pi'(a)/\{1-\pi(a)\}$. In Section~\ref{sec:sens}, we assess the sensitivity of the results to the former issue, repeating all analyses using simulated serological data for the age range $[40,80)$.

\begin{figure}[ht]
\centering
\includegraphics[width=6cm]{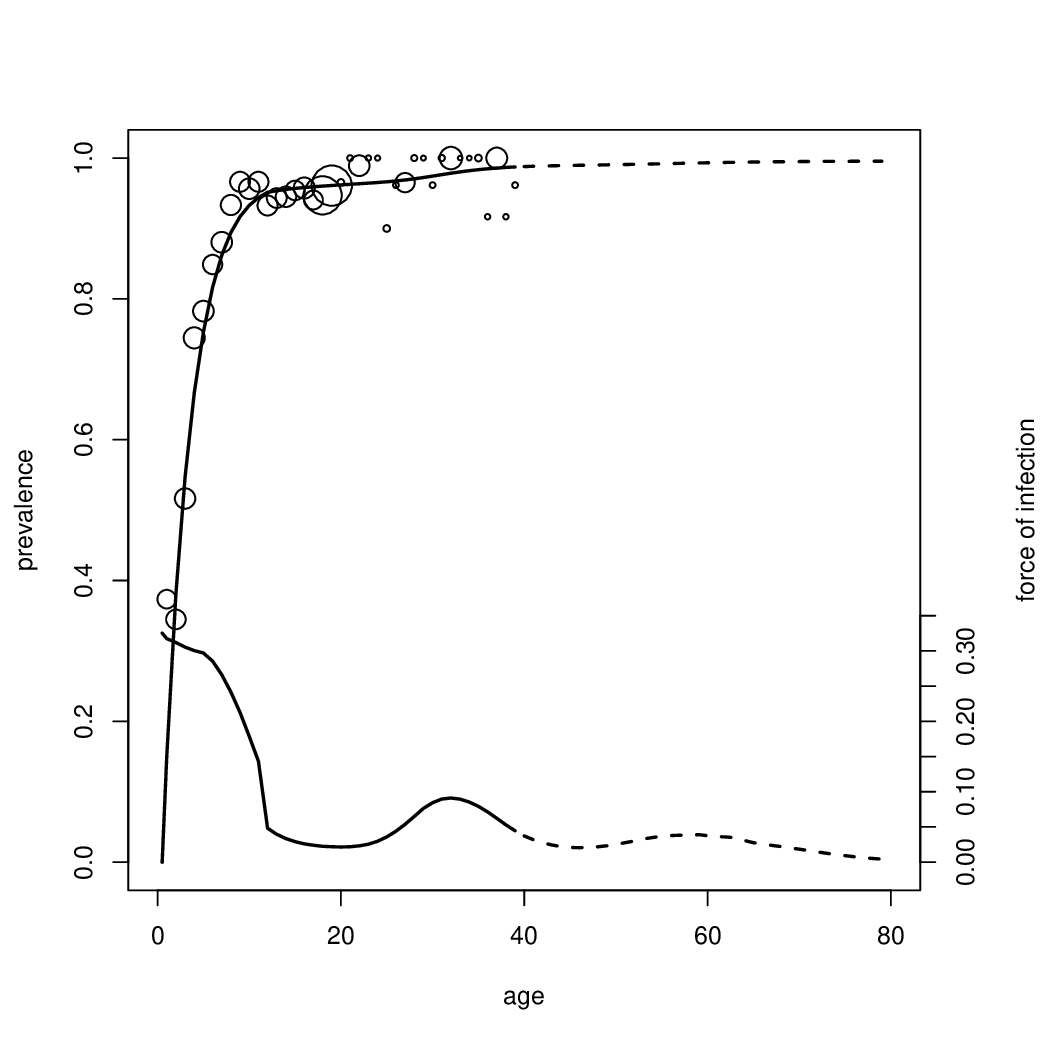} \hspace{1cm}
\includegraphics[width=6cm]{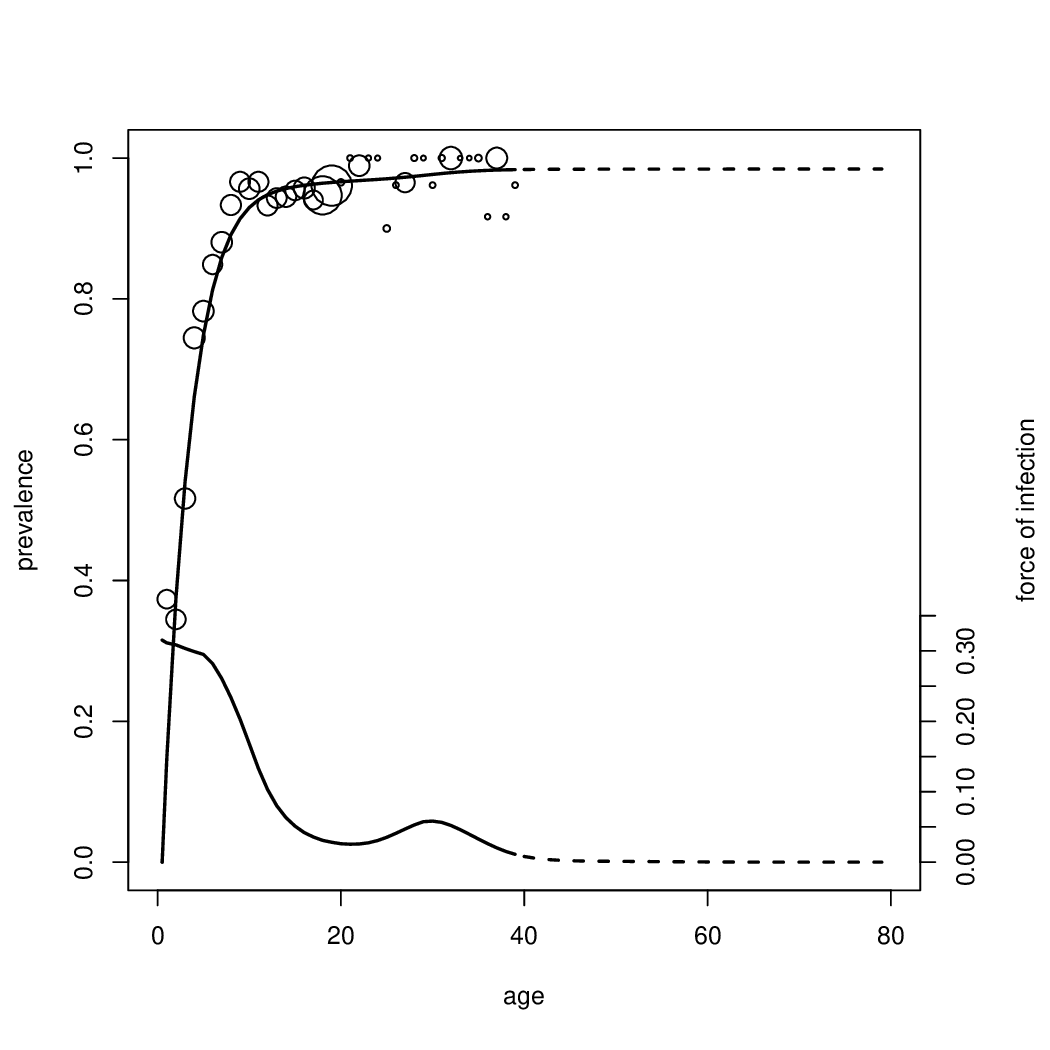}
\caption{Estimated prevalence (upper curve) and force of infection (lower curve) for the discrete model $M_3$ (left) and the continuous model $M_7$ (right).} \label{fig:fitM}
\end{figure}

Figure~\ref{fig:fitM} displays the estimated prevalence function and force of infection for the discrete model $M_3$ (left) and the continuous model $M_7$ (right). The results are remarkably similar. The effect of making $q$ age-dependent is visualized by comparing Figure~\ref{fig:fitM} to the fit of model $C_1$, which was very close to model $C_3$, in Figure~\ref{fig:fitprev} (on the right). The models assuming age-dependent proportionality estimate an initially higher force of infection and a steeper decrease from the age of ten, after which the force of infection is reduced by a factor two, compared to the constant proportionality model. While the latter model predicts total immunity for VZV at older ages, the age-dependent proportionality models estimate a fraction of seropositives which is below one at all times.

\subsection{Model selection and multi-model inference} \label{sec:mminf}

Table~\ref{tab:candidate} presents all candidate models for the proportionality factor $q$ we have collected up till now, among which the constant proportionality model $C_3$, the discrete age-dependent proportionality models $M_1$, $M_2$ and $M_3$, and the continuous age-dependent proportionality models $M_6$, $M_7$ and $M_8$. Further for each model, the number of parameters $K$, the AIC-value, the AIC difference $\Delta_k$, the Akaike weight $w_k$ and the evidence ratio (ER) are displayed.

Model $M_3$ with an assortative component $\gamma_1$ and a background component $\gamma_2$ is the `best' model for $q$ according to the AIC-criterion. However, model selection uncertainty is likely to be high since the selected best model has an Akaike weight of only 0.293 \citep{Burnham2002}. The evidence ratios for $M_3$ versus $M_1$ and $M_2$ are both 1.1, which means there is weak support for the best model. If many independent samples could be drawn, the three discrete age-dependent models would probably compete each other for the `best' model position. The continuous models $M_6$ and $M_7$ have evidence ratios around 3.5, indicating that these models also contribute some information. Models $C_3$ and $M_8$ have the largest AIC difference $\Delta_k$, a very small Akaike weight ($\leq 0.005$) and very large evidence ratios (84.9 and 61.9 respectively), which means there is little support for these two models.

Since there is no single model in the candidate set that is clearly superior to the others and since the estimate for the basic reproduction number $R_0$ varies noticeably over the candidate models, we are not inclined to base prediction only on $M_3$. Applying the concepts of model averaging, as described in \cite{Burnham2002}, a weighted estimate of $R_0$ is calculated, based on the model estimates and the corresponding Akaike weights:
\[\widehat{\overline{R}}_0 = \sum_{k=1}^{7} w_k (\widehat{R}_0)_k = 6.07.\]
With the bootstrap procedure, we obtain a 95\% percentile confidence interval for this model averaged estimate $\widehat{\overline{R}}_0$, namely $[4.4, 351.6]$. Again, there is a large upper limit induced by the same issues reported in Section~\ref{sec:continuous}.

\subsection{Sensitivity analysis} \label{sec:sens}

In order to assess the lack-of-data-problem, we simulate serological data for the age range $[40,80)$ using a constant prevalence $\pi=0.983$, which is estimated from a thin plate regression spline model for the original serological data. Sample sizes for one-year age groups are chosen according to the Belgian population distribution in 2003 \citep{Federale2006} and the total size of serological data now amounts to $n=3856$. The seven candidate models for the proportionality factor $q$ are now applied to the `complete' serological data set.

\begin{figure}[h]
\centering
\includegraphics[width=6cm]{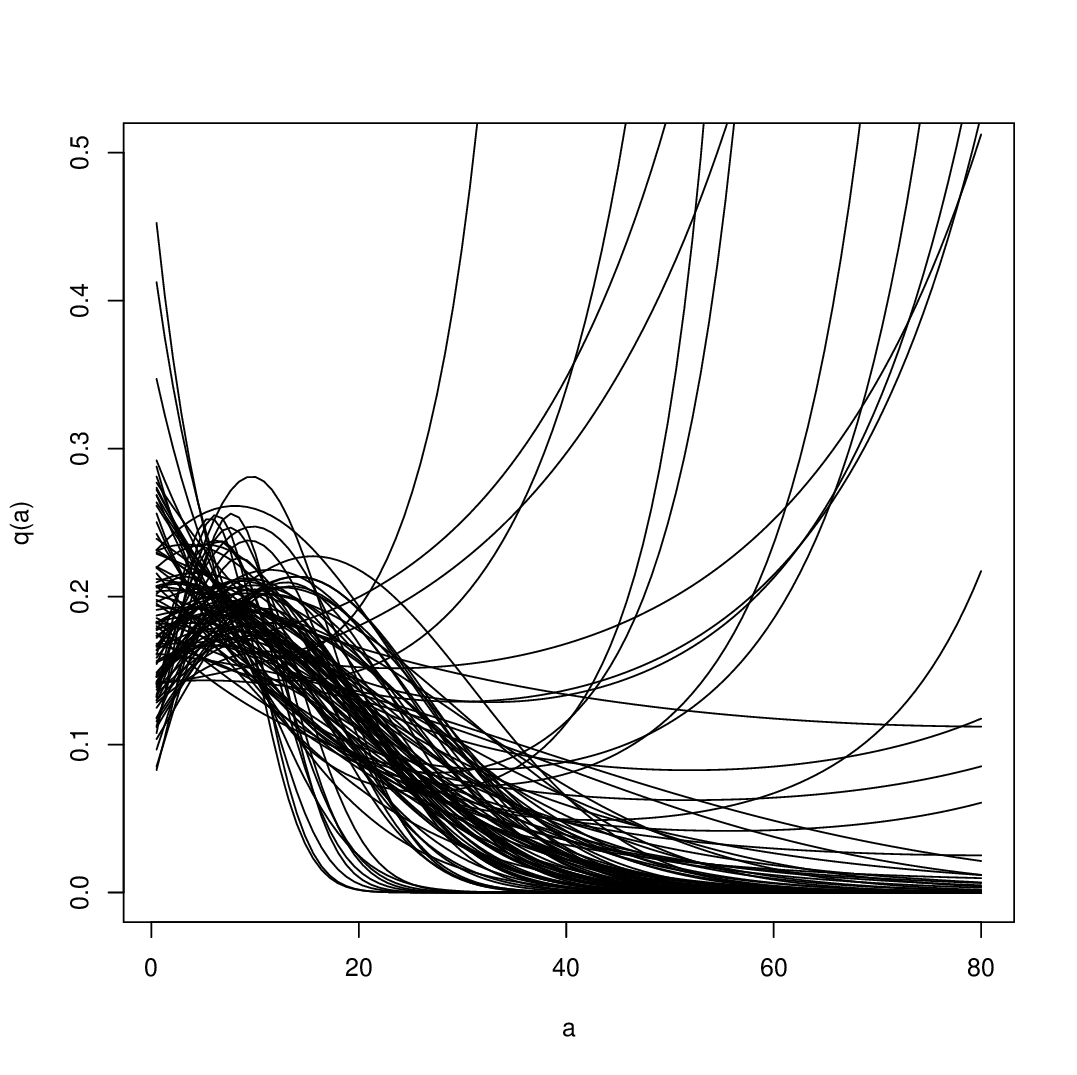} \hspace{1cm}
\includegraphics[width=6cm]{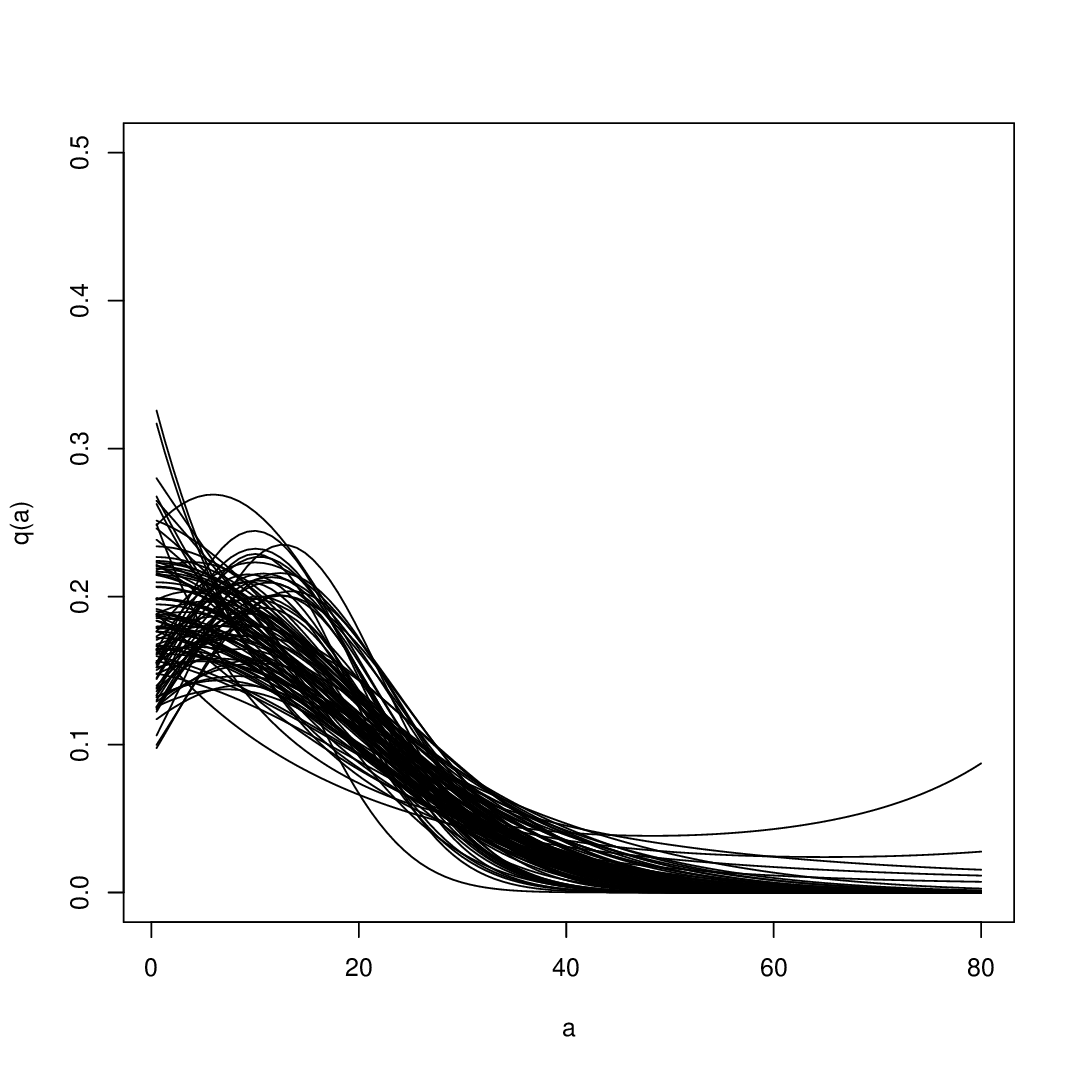}
\caption{$q(a)$ estimates for model $M_7$, shown for 100 randomly chosen bootstrap samples from the original serological data (left) and from this data augmented with simulated data for $[40,80)$ (right).} \label{fig:M7}
\end{figure}

The results are presented in Table~\ref{tab:simulation} and are, overall, quite similar to the results obtained before (Table~\ref{tab:candidate}). The 95\% percentile confidence intervals for $R_0$ ($B=599$ bootstrap samples converged out of 700), however, are narrower since the simulated data for the age range $[40,80)$ are `forcing' the proportionality factor $q$ to follow a natural pace. This is illustrated for model $M_7$ in Figure~\ref{fig:M7}, where the estimated function $q(a)$ is depicted for 100 randomly chosen bootstrap samples. Particularly, right confidence interval limits for $R_0$ are smaller, whereas for most models the $R_0$ estimate seems to have decreased just a little bit.

Model selection uncertainty is illustrated quite nicely here, since four models, $M_7$, $M_3$, $M_2$ and $M_1$, have Akaike weights close to 0.24 and these models also had the most support for the original data set (Table~\ref{tab:candidate}). The model averaged estimate $\widehat{\overline{R}}_0$ now equals 5.64 and the 95\% bootstrap-based  percentile confidence interval is $[4.7, 7.5]$.

\begin{table}
\caption{\label{tab:simulation}Candidate models for the proportionality factor applied to the serological data set augmented with simulated data, together with ML-estimates for the transmission parameters and $R_0$,  95\% bootstrap-based percentile confidence intervals,  and several measures related to model selection.}
\centering\fbox{%
\small
\begin{tabular}{ccrcccccrrr}
Model &  \multicolumn{2}{c}{Parameter} & 95\% CI & $\widehat{R}_0$ & 95\% CI for $R_0$ & $K$ & AIC & $\Delta_k$ & $w_k$ & ER \\
\hline \vspace{-0.2cm} \\
$C_3$ & $\hat{q}$ & 0.159  &  [0.126, 0.195]   & 7.98  &  [6.60, 10.19] & 1 & 1618.747 & 70.774 & $\ll0.0001$ & $\gg10^{3}$ \\
$M_1$ & $\hat{\gamma}_1$ & 0.189 & [0.137, 0.250] &  4.20 & [3.88, 5.74] & 2 & 1548.714 & 0.741 & 0.201 & 1.4  \\
& $\hat{\gamma}_2$ & 0.052 & [0.021, 0.095] & & & & & & &  \\
$M_2$ & $\hat{\gamma}_1$ & 0.186 & [0.136, 0.247] & 4.74 & [4.36, 6.07] & 2 & 1548.627 & 0.654 & 0.210 & 1.4 \\
& $\hat{\gamma}_2$ & 0.052 & [0.020, 0.091] & & &  & & & & \\
$M_3$ & $\hat{\gamma}_1$ & 0.189 & [0.137,  0.250] & 8.28 & [6.43, 11.52] & 2 & 1548.344 & 0.371 & 0.242 & 1.2 \\
& $\hat{\gamma}_2$ & 0.044 & [0.016,  0.082] & & & & & & & \\
$M_6$ &  $\hat{\gamma}_0$ & -1.561 & [-1.934, -1.120] & 4.96 & [4.47, 6.54] & 2 & 1551.321 & 3.348 & 0.055 & 5.3 \\
&  $\hat{\gamma}_1$ & -0.035 & [-0.067, -0.014] & & & & & & & \\
$M_7$ &  $\hat{\gamma}_0$ & -1.793  & [-2.247, -1.079] & 5.22 & [4.60, 7.51] & 3 & 1547.973 & 0 & 0.292 & 1.0 \\
& $\hat{\gamma}_1$ & 0.030 & [-0.074, 0.126] & & & & & & &\\
& $\hat{\gamma}_2$ & -0.002  & [-0.006, 0.001] & & & & & & &\\
$M_8$ &  $\hat{\gamma}_0$ & -1.458 & [-2.061, -0.844] &  2.69 & [2.08, 12.97] & 2 & 1610.113 & 62.140 & $\ll0.0001$ & $\gg10^{3}$ \\
&  $\hat{\gamma}_1$ & -0.103 & [-0.254, 0.016] & & & & & & &\\
\end{tabular}}
\end{table}
\normalsize

\section{Concluding remarks}

In this paper, an overview of different estimation methods for infectious disease parameters from data on social contacts and serological status, was given. The theoretical framework included a compartmental MSIR-model, taking into account the presence of maternal antibodies, and the mass action principle, as presented by \cite{Anderson1991}. An important assumption made was the one of endemic equilibrium, which means that infection dynamics are in a steady state. The serological data set we used was collected over 17~months, averaging over potential epidemic cycles of VZV in Belgium during that period.
In Section~3, we have illustrated the traditional, basic approach of imposing mixing patterns on the WAIFW-matrix to estimate transmission parameters from serological data. In contrast, the novel approach of using social contact data to estimate infectious disease parameters, avoids the choice of a parametric model for the entire WAIFW-matrix.

The idea is fairly simple: transmission rates for infections that are transmitted from person to person in a non-sexual way, such as VZV, are assumed to be proportional to rates of making conversational and/or physical contact, which can be estimated from contact surveys. Although more time consuming, the bivariate smoothing approach as proposed in Section~4, was better able to capture important features of human contacting behavior, compared to the maximum likelihood estimation method of \cite{Wallinga2006}. However, when a non-parametric bootstrap approach was applied to take into account sampling variability, convergence problems arose, probably due to the large number of zeros in combination with the log-link. Therefore, a mixture of Poisson distributions or a zero-inflated negative binomial distribution could be more appropriate. Further, in Section~4, we dealt with a couple of challenges posed by \cite{Halloran2006}. The social contact survey contained useful additional information on the contact itself, which allowed us to target very specific types of contact with high transmission potential for VZV. Furthermore, a non-parametric bootstrap approach was proposed to improve statistical inference.

The constant proportionality assumption was relaxed in Section 5 and we have shown that an improvement of fit could be obtained by disentangling the transmission rates into a product of two age-specific variables: the age-specific contact rate and an age-specific proportionality factor. The latter may reflect, for instance, differences in characteristics related to susceptibility and infectiousness or discrepancies between the social contact proxies measured in the contact survey and the true contact rates underlying infectious disease transmission. We would like to emphasize that there probably exist other models for $q(a,a')$ than the ones considered in Section 5, which fit the data even better.  Our choice of a set of plausible candidate models was directed by parsimony on the one hand, limiting the total number of parameters to three, and prior knowledge on the other hand, considering loglinear models. Furthermore, we restricted analyses to close contacts lasting longer than 15 minutes,  which means that close contacts of short duration and non-close contacts are assumed not to contribute to transmission of VZV.

\begin{figure}[ht]
\centering
\includegraphics[width=13.5cm]{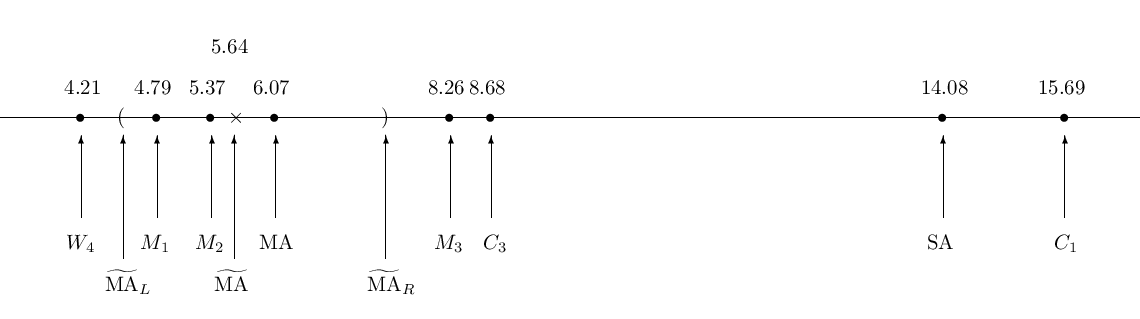}
\caption{$R_0$ estimates for mixing pattern $W_4$, applied to the serological data in Section \ref{sec:vzvappl1}, and for the following models using social contact data: the saturated model (SA) as proposed by \cite{Wallinga2006}, applied in Section \ref{sec:vzvappl2} assuming constant proportionality, and further bivariate smoothing models: constant proportionality models $C_1$ and $C_3$ considering all and close contacts longer than 15 minutes, respectively (Section \ref{sec:refine}) and discrete age-dependent proportionality models $M_1$, $M_2$ and $M_3$ (Section \ref{sec:discrete}). The model averaged estimates for $R_0$ calculated from Table \ref{tab:candidate} (MA), based on the original serological data, and from Table \ref{tab:simulation} ($\widetilde{\mbox{MA}}$), based on the serological data set augmented with simulated data, are displayed, as well as 95\% bootstrap-based percentile confidence interval limits for the latter: $[\widetilde{\mbox{MA}}_L, \widetilde{\mbox{MA}}_R]$.} \label{fig:interval}
\end{figure}

It is important to note that different assumptions concerning the underlying type of contact as well as different parametric models for $q(a,a')$, are likely to entail different estimates of $R_0$, however, they may still induce a similar fit to the serological data. In order to deal with this problem of model selection uncertainty we have turned to multi-model inference in Section~\ref{sec:mminf}. In Figure~\ref{fig:interval}, estimates of $R_0$ are presented for the main estimation methods considered in this paper: the traditional method of imposing mixing patterns to the WAIFW-matrix ($W_4$) and the method of using data on social contacts, assuming constant proportionality (the saturated model SA, $C_1$ and $C_3$) and age-dependent proportionality ($M_1$, $M_2$ and $M_3$). There is a pronounced variability in the estimates of $R_0$, which is partially captured by the model averaged estimate MA, calculated from Table \ref{tab:candidate}.

When estimating $q(a,a')$, we were actually faced with three problems of indeterminacy. First, there is lack of serological information for individuals aged 40 and older, second, prevalence of VZV rapidly stagnates, leading to an indeterminate force of infection and third, serological surveys do not provide information related to infectiousness. Models which only expressed age differences in $q$ for infectious individuals, such as the discrete model $M_5$ (Section~\ref{sec:discrete}) and the continuous models $M_8$ and $M_9$ (Section~\ref{sec:continuous}), either did not lead to convergence or induced unrealistically large bootstrap estimates for $q$ at older ages.

A sensitivity analysis in Section~\ref{sec:sens} showed that lack of serological data had a big impact on confidence intervals for $R_0$. We simulated data for the age range $[40,80)$, giving rise to a model averaged estimate $\widetilde{\mbox{MA}}$ as displayed in Figure~\ref{fig:interval} with corresponding confidence interval limits $[\widetilde{\mbox{MA}}_L, \widetilde{\mbox{MA}}_R]$. The latter problems of indeterminacy might be controlled by combining information on the same infection over different countries or on different airborne infections, assuming there is a relation between the country- or disease-specific $q(a,a')$, respectively. This strategy already appeared beneficial when estimating $R_0$ directly from seroprevalence data, without using social contact data \citep{Farrington2001}.

Further, the impact of intervention strategies such as school closures, might be investigated by incorporating transmission parameters, estimated from data on social contacts and serological status, in an age-time-dynamical setting. Finally, it is important to note that the models rely on the assumptions of type I mortality and type I maternal antibodies in order to facilitate calculations. Consequently, model improvements could be made through a more realistic approach of demographical dynamics.

\section*{Acknowledgements}

This study has been made and funded as part of ``SIMID'', a strategic basic research project funded by the institute for the Promotion of Innovation by Science and Technology in Flanders (IWT), project number 060081. This work has been partly funded and benefited from discussions held in POLYMOD, a European Commission project funded within the Sixth Framework Programme, Contract number: SSP22-CT-2004-502084. We also gratefully acknowledge support from the IAP research network nr P6/03 of the Belgian Government (Belgian Science Policy).

\bibliographystyle{chicago}

\end{document}